\documentstyle[12pt,epsf]{article}
\title{Spin-$\frac{1}{2}$ transverse $XX$ chain
       with a correlated disorder:
       \protect\\
       dynamics of the transverse correlations}
\author{Oleg Derzhko$^{\dagger,\ddagger}$
        and
        Taras Krokhmalskii$^{\dagger}$\\
\small {\em {$^{\dagger}$Institute for Condensed Matter Physics,}}\\
\small {\em {1 Svientsitskii St., L'viv--11, 290011, Ukraine}}\\
\small {\em {$^{\ddagger}$Chair of Theoretical Physics, 
             Ivan Franko State University of L'viv,}}\\
\small {\em {12 Drahomanov St., L'viv--5, 290005, Ukraine}}}

\date{\today}

\begin{document}

\maketitle

\begin{abstract}
We examine numerically the dynamics of $zz$ correlations in the 
spin-$\frac{1}{2}$ isotropic $XY$ chain with random intersite 
coupling and on--site transverse field that depends linearly on 
the neighbouring couplings (correlated off--diagonal and diagonal 
disorder). We discuss the changes in the frequency profiles of $zz$ dynamic 
structure factor caused by disorder.
\end{abstract}

\vspace{1cm}


\vspace{1cm}


\vspace{1mm}

\noindent
{\bf Postal address:}\\
{\em
Dr. Taras Krokhmalskii (corresponding author)\\
Institute for Condensed Matter Physics\\
1 Svientsitskii St., L'viv--11, 290011, Ukraine\\
Tel: (0322) 76 09 08\\
Fax: (0322) 76 19 78\\
E-mail: krokhm@icmp.lviv.ua

\clearpage

\renewcommand\baselinestretch {1.85}
\large\normalsize

Recently the properties of the 
spin-$\frac{1}{2}$ transverse $XX$ chain with correlated disorder
have been discussed in some detail \cite{001,002}.
Such a model consists of
$N\to\infty$ spins $\frac{1}{2}$
on a circle governed by the Hamiltonian
\begin{eqnarray}
\label{001}
H=\sum_{n=1}^{N}\Omega_ns_n^z
+\sum_{n=1}^{N}J_n\left(s^x_ns^x_{n+1}+s^y_ns^y_{n+1}\right).
\end{eqnarray}
It is assumed that the intersite couplings $J_n$ are 
independent random variables
each with the probability distribution
$p(J_n)$
and the on--site transverse field $\Omega_n$
is determined  by the surrounding couplings $J_{n-1}$ and $J_n$
according to the formula
\begin{eqnarray}
\label{002}
\Omega_n=\overline{\Omega}
+\frac{a}{2}\left(J_{n-1}+J_n-2\overline{J}\right)
\end{eqnarray}
where $\overline{\Omega}$ and $\overline{J}$ are the mean values of 
$\Omega_n$ and $J_n$, respectively,
and $a$ is a real parameter.
The models with correlated disorder should arise while describing materials 
with topological disorder. Although there are a few examples of real 
materials which are reasonably well described by the one--dimensional 
spin-$\frac{1}{2}$ isotropic $XY$ model (see, for example, \cite{003}) the 
introduced model (\ref{001}), (\ref{002}) to our best knowledge was not  
related to any particularly compound. However, it is still of much use for 
understanding the generic effects of disorder since in the case of the 
Lorentzian probability distribution
$p(J_n)$
and $\vert a\vert\ge 1$ 
it is possible to find explicitly the exact expression for the 
random--averaged density of states
$\overline{\rho(E)}$ 
($\overline{(\ldots)}
\equiv\ldots\int_{-\infty}^{\infty}dJ_np(J_n)\ldots (\ldots)$)
and thus to examine rigorously the thermodynamic 
properties of a magnetic model with randomness \cite{001}.

The obtained up till now exact analytical results pertain only to 
thermodynamics. In the present paper we study the effects of 
correlated disorder on 
the dynamics of spin correlations examining for this 
purpose the $zz$ dynamic structure factor
\begin{eqnarray}
\label{003}
\overline{S_{zz}(\kappa,\omega)}
=
\int_{-\infty}^{\infty}dt
{\mbox{e}}^{-\epsilon\vert t\vert}
{\mbox{e}}^{i\omega t}
\sum_{n=1}^N{\mbox{e}}^{i\kappa n}
\left[
\overline{\langle s_j^z(t)s_{j+n}^z\rangle}
-\overline{\langle s_j^z\rangle\langle s_{j+n}^z\rangle}
\right],
\;\;\;
\epsilon\rightarrow +0.
\end{eqnarray}
The evaluation of the $zz$ time--dependent spin correlation functions 
$\overline{\langle s_j^z(t)s_{j+n}^z\rangle}$ cannot be performed 
analytically but it can be done numerically \cite{004} 
(see also \cite{005, 006, 007, 008}). 
In what follows we consider the rectangle (but not Lorentzian) probability 
distribution
\begin{eqnarray}
\label{004}
p(J_n)=\frac{1}{2\Delta}\Theta(J_n-\overline{J}+\Delta)
\left[
1-\Theta(J_n-\overline{J}-\Delta)
\right],
\end{eqnarray}
where $\Delta$ controls the strength of disorder. 
From (\ref{002}), (\ref{004}) one can find the probability distribution for 
the random variable $\Omega_n$
\begin{eqnarray}
\label{005}
p(\Omega_n)
=\frac{1}{\vert a\vert\Delta}
\left(
1-\frac{1}{\vert a\vert\Delta}
\vert\Omega_n-\overline{\Omega}\vert
\right)
\nonumber\\
\times
\Theta(\Omega_n-\overline{\Omega}+\vert a\vert\Delta)
\left[
1-\Theta(\Omega_n-\overline{\Omega}-\vert a\vert\Delta)
\right].
\end{eqnarray}
To reveal the effects of correlated disorder besides the model defined by 
(\ref{001}), (\ref{004}), (\ref{002}) we consider the case of 
non--correlated disorder for model (\ref{001}) 
assuming that $J_n$ and $\Omega_n$ 
are independent random variables with probability distributions (\ref{004}) 
and (\ref{005}), respectively. In our computations we considered chains of 
$N=400$ spins with $\overline{J}=-1$, $\overline{\Omega}=0.5$ at low 
temperature $\beta=1000$. We took $\Delta=0.5$ for correlated disorder with 
$a=\pm 1.01$ and for non--correlated disorder and performed the random 
averaging of the $zz$ dynamic correlation functions over $3000$ random 
realizations. We put in (\ref{003}) $j=150$ and computed 
correlation functions with $n$ up to $100$ 
for the times up to $15,\ldots,160$ (depending on the value of $\kappa$). 
We adopted $\epsilon=0.001$. To prove that our results for the taken 
values of parameters already pertain to thermodynamic systems 
we performed many 
additional calculations similar to that described in \cite{004}. 
The main results of our study are shown in Fig.\ \ref{fig1}
\begin{figure}[t]
\epsfysize=34mm
\epsfclipon
\centerline{\epsffile{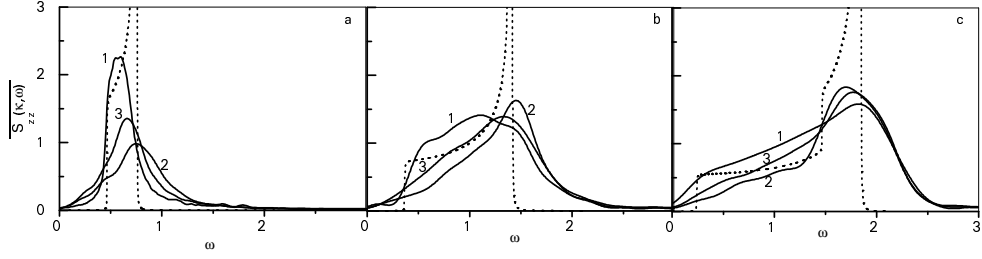}}
\caption[ ]{Frequency dependence of 
 $\overline{S_{zz}(\kappa,\omega)}$ 
 at 
 $\kappa=\frac{\pi}{4}$ (a),
 $\kappa=\frac{\pi}{2}$ (b),
 $\kappa=\frac{3\pi}{4}$ (c)
 at low temperature $\beta=1000$ 
 for model (\ref{001}) with $\overline{J}=-1$, $\overline{\Omega}=0.5$,
 $\Delta=0.5$;
 1 --- correlated disorder with $a=-1.01$,
 2 --- correlated disorder with $a=1.01$,
 3 --- non--correlated disorder.
 Dashed curves correspond to the non--random case.}
 \label{fig1}
\end{figure}
where we displayed the frequency dependence of 
$\overline{S_{zz}(\kappa,\omega)}$ at 
$\kappa=\frac{\pi}{4},\;\frac{\pi}{2},\;\frac{3\pi}{4}$ for different types 
of disorder.

Let us turn to a discussion of the obtained results. Dynamics of the 
transverse correlations in the non--random case is well known 
\cite{009,010,011}. In 
the Jordan--Wigner picture the zero--temperature $zz$ dynamic properties of 
the spin-$\frac{1}{2}$ transverse $XX$ chain are conditioned by exciting of 
two fermions with energies 
$\Lambda_{\kappa^{\prime}}=\Omega+J\cos\kappa^{\prime}<0$
and 
$\Lambda_{\kappa^{\prime\prime}}=\Omega+J\cos\kappa^{\prime\prime}>0$,
for which
$\omega=-\Lambda_{\kappa^{\prime}}+\Lambda_{\kappa^{\prime\prime}}$
and
$\kappa^{\prime\prime}=\kappa^{\prime}-\kappa$.
Consider, for example, 
$S_{zz}(\frac{\pi}{4},\omega)$
(dashed curve in Fig.\ \ref{fig1}a).
Evidently,  
$-\frac{\pi}{3}<\kappa^{\prime}<-\frac{\pi}{12}$, 
$-\Lambda_{\kappa^{\prime}}+\Lambda_{\kappa^{\prime}-\frac{\pi}{4}}
=-2\sin\frac{\pi}{8}\sin\left(\kappa^{\prime}-\frac{\pi}{8}\right)$
and hence the lower frequency at which 
$S_{zz}(\frac{\pi}{4},\omega)$ 
appears is equal to $\approx 0.466$
(two fermions with the energies 
$\Lambda_{-\frac{\pi}{12}}\approx -0.466$ 
and 
$\Lambda_{-\frac{\pi}{3}}=0$, 
respectively),
the upper frequency after which 
$S_{zz}(\frac{\pi}{4},\omega)$ 
disappears is equal to $\approx 0.759$
(two fermions with the energies 
$\Lambda_{-\frac{\pi}{3}}=0$ 
and 
$\Lambda_{-\frac{7\pi}{12}}\approx 0.759$, 
respectively). 
We may relate the changes in the transverse dynamic structure factor 
due to randomness 
to the changes in the random--averaged density of states 
for different types of disorder (Fig.\ \ref{fig2}).
Indeed, the pair of fermions determining the lower frequency 
roughly speaking 
does exist 
for $a=-1.01$ and does not exist for $a=1.01$ and for non--correlated 
disorder, whereas the density of states for the energies corresponding 
to the pair of fermions determining the upper frequency is diminished 
equally because of disorder in all three cases. 

Consider further $S_{zz}(\frac{\pi}{2},\omega)$.
Repeating the above arguments one concludes that 
two fermions with the energies 
$\Lambda_{\frac{\pi}{6}}\approx -0.366$ 
and 
$\Lambda_{-\frac{\pi}{3}}=0$ 
determine the lower frequency $\approx 0.366$,
starting from the frequency $\approx 1.366$ two pairs of fermions 
contribute to the transverse dynamic properties 
(e.g., at that frequency one finds 
two fermions with the energies 
$\Lambda_{-\frac{\pi}{6}}\approx -0.366$ 
and 
$\Lambda_{-\frac{2\pi}{3}}=1$ 
and another pair of
fermions with the energies 
$\Lambda_{-\frac{\pi}{3}}=0$ 
and 
$\Lambda_{-\frac{5\pi}{6}}\approx 1.366$), 
and two fermions with the energies 
$\Lambda_{-\frac{\pi}{4}}\approx -0.207$ 
and 
$\Lambda_{-\frac{3\pi}{4}}\approx 1.207$ 
determine the upper frequency $\approx 1.414$.
Analysing $\overline{\rho(E)}$ in Fig.\ \ref{fig2} 
one observes, for example, that 
$\overline{\rho(E)}$ at the energies related to the lower frequency is 
almost not diminished for the correlated disorder
with $a=-1.01$, 
whereas $\overline{\rho(E)}$ is diminished essentially for 
$a=1.01$ and
non--correlated disorder. This is in agreement with the changes in 
$\overline{S_{zz}(\frac{\pi}{2},\omega)}$ 
seen in Fig.\ \ref{fig1}b.

Consider finally $S_{zz}(\frac{3\pi}{4},\omega)$.
Similarly to the previous cases one finds that the lower frequency 
$\approx 0.241$ 
is conditioned by two fermions 
with the energies 
$\Lambda_{\frac{\pi}{3}}=0$ 
and 
$\Lambda_{-\frac{5\pi}{12}}\approx 0.241$,
starting from the frequency $\approx 1.466$
$S_{zz}(\frac{3\pi}{4},\omega)$ is determined by two pairs of fermions
(e.g., at that frequency one finds
two fermions with the energies 
$\Lambda_{\frac{\pi}{12}}\approx -0.466$ 
and 
$\Lambda_{-\frac{2\pi}{3}}=1$
and 
another pair of fermions with the energies 
$\Lambda_{-\frac{\pi}{3}}=0$ 
and 
$\Lambda_{-\frac{13\pi}{12}}\approx 1.466$),
and the upper frequency 
$\approx 1.848$ 
is conditioned by two fermions 
with the energies 
$\Lambda_{-\frac{\pi}{8}}\approx -0.424$ 
and 
$\Lambda_{-\frac{7\pi}{8}}\approx 1.424$.
From Fig.\ \ref{fig2} one notes that in the case of lower frequency the 
disorder decreases $\overline{\rho(E)}$ at the corresponding energies for 
non--correlated disorder and the correlated one with 
$a=1.01$ stronger than for 
the correlated disorder with
$a=-1.01$, whereas 
in the case of the upper frequency $\overline{\rho(E)}$ 
at the corresponding energies is diminished more for non--correlated 
disorder than for correlated disorder that is consistent with frequency 
profiles seen in Fig.\ \ref{fig1}c.
\begin{figure}[t]
\epsfysize=34mm
\epsfclipon
\centerline{\epsffile{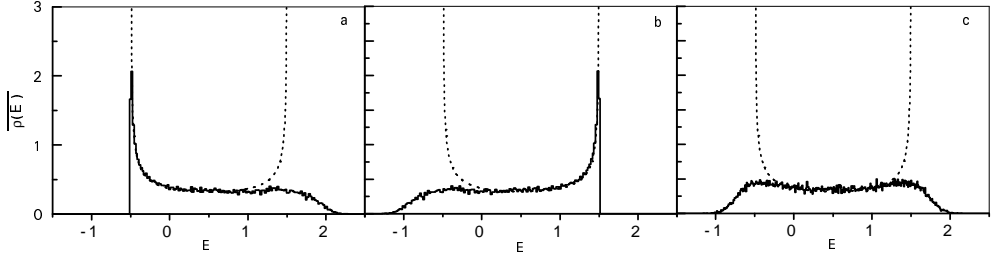}}
 \caption[ ]{$\overline{\rho(E)}$ 
 evaluated numerically (see \cite{004})
 for model (\ref{001}) with $\overline{J}=-1$, $\overline{\Omega}=0.5$,
 $\Delta=0.5$;
 a --- correlated disorder with $a=-1.01$,
 b --- correlated disorder with $a=1.01$,
 c --- non--correlated disorder.
 Dashed curves correspond to the non--random case.}
 \label{fig2}
\end{figure}

To summarize, we examined the low--temperature dynamics of the transverse 
spin correlations in the spin-$\frac{1}{2}$ transverse $XX$ chain with 
correlated disorder computing the transverse dynamic structure factor 
$\overline{S_{zz}(\kappa,\omega)}$. 
We found that within certain frequency regions the 
introducing of disorder may yield almost no changes in the 
value of $\overline{S_{zz}(\kappa,\omega)}$ 
(e.g., $\overline{S_{zz}(\frac{\pi}{4},\omega)}$
at $\omega\approx 0.5$ for $a=-1.01$
or $\overline{S_{zz}(\frac{\pi}{2},\omega)}$
at $\omega\approx 1$ for non--correlated disorder).
We observed that the changes in 
$\overline{S_{zz}(\kappa,\omega)}$ caused by disorder may be explained by 
the changes in the random--averaged density of states. 
Evidently, because of the Jordan--Wigner mapping the obtained results may be 
useful for understanding the conductivity in a chain of tight--binding 
fermions with random correlated hopping and on--site energy.

\vspace*{0.25cm} 
The authors are grateful to Prof. M. Shovgenyuk for providing a possibility 
to perform numerical calculations.
They are indebted to the
organizers 
of the International Conference
LOCALIZATION 1999
(Hamburg, 1999)
for the financial support for attending the conference.
%
%
%
%
%
%
%
%
%
%
%
%

\end{document}